\documentclass[apj]{emulateapj}
\usepackage{color}
\usepackage{natbib}

\shorttitle{Assembly history of massive galaxies}
\shortauthors{Jaehyun Lee and Sukyoung K. Yi}

\begin{document}
\title{On the assembly history of stellar components in massive galaxies}
\author{Jaehyun Lee and Sukyoung K. Yi}
\affil{Department of Astronomy and Yonsei University Observatory, Yonsei University, Seoul 120-749, Republic of Korea; yi@yonsei.ac.kr}

\begin{abstract}

~\citet{matsuoka10} showed that the number density of the most massive galaxies ($\log M/M_{\odot}=11.5-12.0$) increases faster than that of the next massive group ($\log M/M_{\odot}=11.0-11.5$) during  $0 < z < 1$. This appears to be in contradiction to the apparent ``downsizing effect''. We attempt to understand the two observational findings in the context of the hierarchical merger paradigm using semi-analytic techniques. Our models closely reproduce the result of~\citet{matsuoka10}. Downsizing can also be understood as larger galaxies have, on average, smaller assembly ages but larger stellar ages. Our fiducial models further reveal  details of the history of the stellar mass growth of massive galaxies. The most massive galaxies ($\log M/M_{\odot}=11.5-12.0$ at z=0), which are mostly brightest cluster galaxies, obtain roughly 70\% of their stellar components via merger accretion. The role of merger accretion monotonically declines with galaxy mass:  40\% for $\log M/M_{\odot}=11.0-11.5$ and 20\% for $\log M/M_{\odot}=10.5-11.0$ at $z=0$. The specific accreted stellar mass rates via galaxy mergers decline very slowly during the whole redshift range, while specific star formation rates sharply decrease with time. In the case of the most massive galaxies, merger accretion becomes the most important channel for the stellar mass growth at $z\sim2$. On the other hand, {\em in-situ} star formation is always the dominant channel in $L_*$ galaxies.

\end{abstract}

\section{Introduction}

Dynamical realizations based on the concordance $\Lambda$CDM cosmology~\citep[e.g.][]{spergel07} have been remarkably successful at reproducing large-scale structures in the universe~\citep[e.g.][]{springel06}. In the common perception of this paradigm, large galaxies grow hierarchically through numerous galaxy mergers that follow mergers between dark halos.

Supporting this view, galaxies with disturbed features have frequently been witnessed~\citep[e.g.][]{arp66,schweizer88}. With the advent of deep and wide field surveys, the arguments could be investigated in greater detail. Recent studies based on ultra-deep imaging data are particularly noteworthy. 
From the $\mu_{\rm r} = 28$ mag arcsec$^{-2}$ deep images, ~\citet{vandokkum05} found that about 50$\%$ of red bulge-dominant galaxies in field environments show tidal debris. Merger galaxy fraction has been found to be almost as high in cluster environments ~\citep{sheen12}. ~\citet{kaviraj07} and~\citet{kaviraj10} claimed that residual star formation found in a large fraction ($\sim 30$\%) of local massive early-type galaxies is related to mergers or interactions. As observation techniques reach deeper and hidden nature of galaxies, ``peculiar'' is no longer a synonym of ``rare''. 

High redshift surveys allow us direct investigation of the role of galaxy mergers and interactions on galaxy evolution. Using the K-band Hubble diagram,~\citet{aragon-salamanca98} found that brightest cluster galaxies(BCGs) have increased their mass by a factor of two to four with no or negative luminosity evolution during $0<z<1$. From the GOODS fields,~\citet{bundy09} found that the pair fraction of massive ($\rm log M/M_{\odot} > 11.0$) red spheroidal galaxies is higher than that of less massive systems ($\rm log M/M_{\odot} \sim 10.0$). In addition, the pair fraction unaccompanied by star formation increases with time. They concluded that massive galaxies grow primarily through dry or minor mergers, at least at $z \lesssim 1$. Almost simultaneously, from the UKIDSS and the SDSS II Supernova Survey, ~\citet[][hereafter MK10]{matsuoka10} found that, during $z<1$ the number density of the most massive galaxies ($\log M/M_{\odot}=11.5-12.0$) increased more rapidly than that of the next massive group ($\log M/M_{\odot}=11.0-11.5$).
They also showed that more massive systems have a lower blue galaxy fraction than less massive systems, and that the fraction decreases with time. All these observations seem to imply that, during $z \lesssim 1$, massive galaxies are mainly brought up via mergers.

Not all observations naively support the hierarchical merger picture. The tight color-magnitude relation found among early-type galaxies is more simply, albeit not exclusively, explained by monolithic formation scenarios~\citep[e.g.][]{bower92,kodama97}. Furthermore, the ``downsizing'' effect, where larger galaxies appear to be older and thus suspected to have formed earlier, seems to be inconsistent with the new paradigm~\citep{cowie96,glazebrook04,cimatti04}. On the face of it, the inconsistency seems a counter-evidence of the hierarchical picture; however, some studies have pointed out that downsizing can be understood by the hierarchical paradigm reasonably enough. Based on semi-analytic models,~\citet{delucia06} showed that the star formation rates of all the progenitors of more massive galaxies peak earlier and decrease faster than those of less massive galaxies. In fact, downsizing could be a natural result of the hierarchical clusterings of dark halos.~\citet{neistein06} demonstrated that downsizing appears in the {\em total (combined)} mass evolution of all the progenitors of a dark matter halo. If the stellar mass of a galaxy correlates with the mass of its dark matter halo~\citep{moster10}, downsizing may very well originate from the bottom-up assembly of dark matter halos. Using semi-analytic approaches, many studies have shown that the assembly and formation history of stellar components, especially in massive, red and luminous, or elliptical galaxies, can be different in the hierarchical Universe~\citep{kauffmann96,baugh96,delucia06,delucia07,almeida08}. More direct hydrodynamic simulations performed for smaller volumes have yielded consistent results~\citep{oser10,lackner12}. 

In this study, we look further into the details of the assembly history of stellar components in massive galaxies. We use semi-analytic approaches because these are more effective for constructing models of a large volume of the universe. Motivated mainly by the empirical results of MK10, we investigate the factors that drive the growth of galaxy stellar mass as a function of time and mass.

\section{Model}

We have developed our own semi-analytic model for galaxy formation and evolution. In this section, we briefly introduce physical ingredients, with more focus on the prescriptions that play important roles in this study.

\subsection{Dark Matter Halo Merger Trees and Galaxy Mergers}

In the context of the two-step galaxy formation theory~\citep{white78}, the first step involves the construction of dark halo merger trees. We ran dark matter N-body volume simulations using the GADGET-2 code~\citep{springel05}. The simulation was performed using the standard $\Lambda$CDM cosmology parameters derived from the WMAP 7-year observations, $\Omega_{m} = 0.266$, $\Omega_{\Lambda} = 0.734$, $\sigma_{8} = 0.801$, $\Omega_{\rm b}=0.0449$, and H$_{0} = 71$ km s$^{-1}$ Mpc$^{-1}$~\citep{jarosik11}. The periodic cube size of the simulation was 70$h^{-1}$Mpc on the side with $512^{3}$ collisionless particles, mass of a particle was 1.9$\times10^{8}h^{-1}M_{\odot}$, and mass resolution of a halo was $\sim 10^{10}h^{-1}M_{\odot}$ . We identify halo structures in the simulation box, using a halo finding code developed by~\citet{tweed09}, which is based on the AdaptaHOP~\citep{aubert04}. Halo merger trees were generated by backtracing the infall histories of subhalos with increasing redshift. 
We calculate galaxy merger timescales, using the positional information of subhalos extracted from our N-body simulation. In this study, we assume that the mass distribution of dark matter halos basically follows the Navarro-Frank-White (NFW) profile~\citep{navarro96}.

Subhalos can disappear before arriving at central regions if they are heavily embedded in their host halo density profiles or their mass decreases below the resolution limit of our numerical simulation. We take these numerical artifacts into consideration. Thus, once a halo enters into a more massive halo, we additionally calculate its merger timescale, $t_{\rm merge}$, using the following fitting formula suggested by~\citet{jiang08}:
\begin{eqnarray}\nonumber
t_{\rm merge} {\rm (Gyr)}&=& \frac{0.94\epsilon^{0.60}+  0.60}{2C}  \frac{M_{\rm host}}{M_{\rm sat}}\\  
&&\frac{1}{\ln [1+(M_{\rm host}/M_{\rm sat})]} \frac{R_{\rm vir}}{V_{c}},
\label{eqn:jiang}
\end{eqnarray}
where $\epsilon$ is the circularity of the orbit of a satellite halo, $C$ is a constant, approximately equal to 0.43, $M_{\rm host}$ is the mass of a host halo, $M_{\rm sat}$ is the mass of a satellite halo, $R_{\rm vir}$ is the virial radius of a host halo, and $V_{c}$ is the circular velocity of a host halo at $R_{\rm vir}$.
If a subhalo disappears within $0.1R_{\rm vir}$ of its host halo, a galaxy in the subhalo is regarded as having merged with its central galaxy. On the other hand, if it disappears outside $0.1R_{\rm vir}$, we assume that the galaxy in the subhalo merges with its central galaxy at $t_{\rm merge}$ given by Eq.~\ref{eqn:jiang} after the subhalo becomes a satellite of its host halo. In that case, we consider dynamical friction to analytically compute the positions and velocities of subhalos. We adopt the dynamical friction prescription introduced by~\citet{binney08}:
\begin{eqnarray}\nonumber
%\begin{aligned}
&& \frac {{\rm d}\vec{v}}{{\rm d}t}_{\rm dynf} = -\frac{GM_{\rm sat}(t)}{r^{2}} {\rm ln}\Lambda \left( \frac{V_{c}}{v} \right)^{2} \\ 
                                                           && \left\{ {\rm erf} \left( \frac{v}{V_{c}}\right) - \frac{\sqrt{\pi}}{2} \left( \frac{v}{V_{c}} \right) {\rm exp} \left[ - \left( \frac{v}{V_{c}} \right) ^{2}\right] \right\} \vec{e_{v}},
%\end {aligned}
\end{eqnarray}
\label{eqn: dynf}
where $M_{\rm sat}$ is the mass of a subhalo, which is initially defined as the mass at a previous time step after which the subhalo cannot be resolved in the N-body simulation anymore, $r$ is the distance between the subhalo and the center of its host halo, $\ln \Lambda$ is a Coulomb logarithm with $\Lambda=1+M_{\rm host}/M_{\rm sat}$ adopted by~\citet{springel01}, $V_{c}$ is the circular velocity of the host halo at the virial radius, and $v$ is the orbital velocity of the subhalo. 

During the orbital motion of a satellite halo in its host halo, the dark matter of the satellite halo is stripped due to dynamical friction. If the satellite is resolved in an N-body simulation, dark matter of the satellite would be naturally stripped. On the other hand, if the halo is not resolved in the simulation but considered to orbit around its host halo, stripping due to dynamical friction should be computed analytically. We evaluate the amount of dark matter stripped by dynamical friction by adopting the concept of the sphere of influence ($r_{\rm soi}$) within which dark matter particles are bound to the satellite halos, using the following formula~\citep{battin87}:
\scriptsize
\begin{eqnarray}
%\begin{aligned}
r_{\rm soi} \sim r  \left[ \left( \frac{M_{\rm sat, tot}}{M_{\rm host} (<r)} \right)^{-0.4}(1+3\cos ^2 \theta)^{0.1}+0.4\cos \theta \left( \frac{1+6\cos ^{2} \theta}{1+3\cos ^{2} \theta}\right) \right]^{-1},
%\end {aligned}
\end{eqnarray}
\normalsize
where $r$ is the distance between the centers of the satellite and its host halos, $M_{\rm sat, tot}$ is the total (baryon+dark matter) mass of the satellite halo, $M_{\rm host}(<r)$ is the total mass of the central halo within $r$, and $\theta$ is the angle between the line connecting the particle to the center of the satellite halo and the line connecting the centers of the satellite and the host halos. During a time step, $\delta t$, we assume that a satellite halo loses $\delta M_{\rm sat}=M_{\rm sat}(r>r_{\rm soi})\delta t/t_{\rm dyn}$, where  $t_{\rm dyn}$ is the dynamical timescale of the satellite halo, and $M_{\rm sat}(r>r_{\rm soi})$ is the mass of dark matter outside $r_{\rm soi}$ of the satellite halo.

We assume that stellar components in satellite\footnote{In this paper, galaxies that are not the central one in a halo are all ``satellite''. Only one galaxy is qualified as the central galaxy of a halo and all the rest, regardless of brightness, are satellites.} galaxies merging with their hosts constitute the bulge component of the host. If the mass ratio of baryonic mass ($m_{\rm cold}+m_{*}$) between merging galaxies, $m_{\rm secondary}/m_{\rm primary}$, is greater than 0.25, then it is assumed that all the stellar components of the host galaxy quickly become bulge components of the remnant, as well. 

Empirical studies have shown that intra-cluster light originates from the extended diffuse stellar components of the brightest galaxies in groups or clusters~\citep[e.g.][]{feldmeier02,gonzalez05,zibetti05}. They suggested that diffuse stellar components are the stars scattered from satellite galaxies during tidal stripping or mergers into central galaxies. It has been suggested that 10-40\% of stellar components in satellite galaxies turn into diffuse stellar components in each galaxy merger~\citep{murante04,monaco06}. We adopt the value of 40\% in this study because it resulted in the best reproduction of empirical data. 
The amount of stellar mass that a central galaxy acquires via merger is $(1-f_{\rm scatter})M_{\rm *,sat}$, where $f_{\rm scatter}$ is the fraction of scattered stellar components and $M_{\rm *,sat}$ is the stellar mass of a satellite galaxy.

\subsection{Gas Cooling, Star Formation, and Recycling}
We assume that the baryonic fraction in accreted dark matter follows the global baryonic fraction, $\Omega_{\rm b}/\Omega_{\rm m}$. Baryons are accreted onto dark halos and shock-heated to become hot gas components. %We assume that hot gas follows a sigular isothermal profile truncated at virial radius. Therefore, when a halo undergoes stripping of dark matter, we assume that hot gas is stripped off proportionally.

Gas accretion onto a galactic disk plane via atomic cooling of hot gas is calculated based on the model proposed by~\citet{white91}. The cooling timescale at distance $r$ from the center of a halo is estimated by the following formula:
\begin{eqnarray}
t_{cool}(r)=\frac {3}{2} \frac{\rho_{g}(r) kT}{\mu(Z,T)m_{P}n_{e}n_{i} \Lambda(Z,T)} ,
\label{eqn:t_cool}
\end{eqnarray}
where $\rho_{g}$ is the gas density within radius $r$ of a dark matter halo, $T$ is the temperature of gas, assumed to be the virial temperature in the model, $Z$ is metallicity, $\mu(Z,T)$ is the mean molecular weight of gas with $Z$ and $T$, $m_{P}$ is the mass of a proton, $n_{e}$ is the number density of electrons, $n_{i}$ is the number density of ions, and $\Lambda(Z,T)$ is the cooling function with $Z$ and $T$. The values of $\mu(Z,T)$, $n_{e}$, $n_{i}$, and $\Lambda(Z,T)$ are determined by referring to~\citet{sutherland93}. We do not include self-consistent chemical evolution in our calculation. Instead, we take the metallicity of hot gas components in halos as a constant, $0.3Z_{\odot}$, which is comparable to the metallicity of observed clusters with various masses~\citep{arnaud92}. Because the cooling function, $\Lambda(Z,T)$, is sensitive to metallicity, our results should not be taken too literally. However, relative analysis, a main tool of this study, is affected little by the details in the treatment of chemical evolution.

It is assumed that the gas density follows a singular isothermal profile truncated at $R_{\rm vir}$: $\rho_{g}(r)=m_{\rm hot}/(4\pi R_{\rm vir} r^{2})$. Substituting the formula for $\rho_{g}(r)$ in Eq.~\ref {eqn:t_cool} and adopting the dynamical timescale of a halo, $t_{\rm dyn}$, as $t_{\rm cool}$, one can derive the cooling radius, $r_{\rm cool}$, within which hot gas can cool within $t_{\rm cool}$. For the case of $r_{\rm cool}>R_{\rm vir}$, the cooling rate is rather restrained by the free-fall rate than the cooling rate, so that $\dot{m}_{\rm cool}=m_{\rm hot}/(2t_{\rm cool}$). In contrast, if $r_{\rm cool}<R_{\rm vir}$, $\dot{m}_{\rm cool}=m_{\rm hot}r_{\rm cool}/(2R_{\rm vir}t_{\rm cool}$).

In our model, stars can be formed through a quiescent mode, in which cold gas turns into disk stellar components via gas contraction on a disk, or a burst mode, which is induced by galaxy mergers. Star formation rate in the quiescent mode are delineated by a simple law proposed by~\citet{kauffmann93} as follows:
\begin{eqnarray}
 \dot{m}_{*}=\alpha \frac{m_{\rm cold}}{t_{\rm dyn,gal}},
\label{eqn:quiescent}
\end{eqnarray}
where $\alpha$ is the empirically-determined star formation efficiency, $m_{\rm cold}$ is the amount of cold gas, and $t_{\rm dyn,gal}$ is the dynamical timescale of cold gas disk assumed to be $0.1t_{\rm dyn}$. 

Observations~\citep[e.g.][]{borne00,woods07} and hydrodynamic simulations of galaxy mergers~\citep[e.g.][hereafter C08]{cox08} have shown that galaxy mergers can give rise to rapid star formation. We follow the conventional treatment: stars formed in the quiescent mode belong to a galactic disk, while stars born in the burst mode become bulge components. We adopt the prescription for merger-induced starbursts described in~\citet[][hereafter S08]{somerville08}, which formulates the prescription based on C08. 
 S08 defines burst efficiency, $e_{\rm burst}$,  to parameterize the fraction of the cold gas reservoir involved in a merger induced starburst as follows:
\begin{eqnarray}
\centering
e_{\rm burst}=e_{\rm burst,0} \mu^{\gamma_{\rm burst}},
\end{eqnarray}
where $\mu$ is the mass ratio between a host galaxy and its merger counterpart, $\gamma_{\rm burst}$ is the bulge-to-total mass ratio(B/T) of the host galaxy, and $e_{\rm burst,0}$ is the burst efficiency fitted by the following formula:
\begin{eqnarray}\nonumber
\centering
e_{\rm burst,0}=0.60[V_{\rm vir}/&&({\rm km\,s^{-1}})]^{0.07}(1+q_{\rm EOS})^{-0.17}\\
	&&(1+f_{\rm g})^{0.07}(1+z)^{0.04},
\end{eqnarray}
where $V_{\rm vir}$ is virial velocity, $q_{\rm EOS}$ is the effective equation of state of gas, $f_{\rm g} \equiv m_{\rm cold}/(m_{\rm cold}+m_{*})$ is the fraction of cold gas, and $z$ is the redshift when the disks of progenitor galaxies are constructed. $q_{\rm EOS}$ was suggested to parameterize the multiphase nature of ISM: $q_{\rm EOS}=0$ indicates an isothermal state, and $q_{\rm EOS}=1$ represents the fully pressurized multiphase ISM. In this study, we adopt $q_{\rm EOS}=1$ at which gas is dynamically stable, so that starbursts are more suppressed than the case of $q_{\rm EOS}=0$. The redshift dependency of $e_{\rm burst,0}$ is very weak, and thus we assume $(1+z)^{0.04} \sim 1$. The burst timescale,$\tau_{\rm burst}$, is also formulated by S08 as follows:
\begin{eqnarray}\nonumber
\centering
\tau_{\rm burst}=191{\rm Gyr}[V_{\rm vir}/&&({\rm km\,s^{-1}})]^{-1.88}(1+q_{\rm EOS})^{2.58}\\
	&&(1+f_{\rm g})^{-0.74}(1+z)^{-0.16}.
\end{eqnarray}
The mass ratio, $\mu$, is the ratio of the total mass of central regions ($m_{\rm DM,core}+m_{*}+m_{\rm cold}$) of a host galaxy to that of its merger counterpart. Following S08, we calculate the core mass of a dark matter halo, $m_{\rm DM,core}=m_{\rm DM}(r<2r_{\rm s})$, where $r_{s}\equiv R_{\rm vir}/c_{\rm NFW}$. The concentration index of the Navarro-Frenk-White profile, $c_{\rm NFW}$, is derived based on the fitting function suggested by~\citet{maccio07}.

The parameter $\gamma_{\rm burst}$ is determined by the B/T of a host galaxy as follows:
\begin{eqnarray}
\centering
\gamma_{\rm burst} = \left\{ 
\begin{array}{l l}
	0.61 & \quad \mbox{B/T $\leq$ 0.085}\\
	0.74 & \quad \mbox{0.085 $<$ B/T$\leq$ 0.25}\\
	1.02 & \quad \mbox{0.25 $<$ B/T} \end{array} \right. \
\end{eqnarray}
C08 showed that the burst efficiency of a host galaxy with a high B/T is lower than that of a galaxy with a lower B/T, because more massive bulges stabilize the galaxies and reduce the burst efficiency more effectively. Because C08 demonstrated that mergers with mass ratios below 1:10 are not associated with starbursts, we assume $e_{\rm burst}=0$ if $\mu <0.1$. With the ingredients, the amount of stars born in burst modes is calculated as $m_{\rm burst}=e_{\rm burst}m_{\rm cold}$. We assume that $m_{\rm burst}$ turns into stars for $\tau_{\rm burst}$ in a uniform rate, $\dot{m}_{\rm burst}=m_{\rm burst}/\tau_{\rm burst}$. While merger-induced starbursts occur in a galaxy, the quiescent mode also still goes on in our models. 

In our model, the recycling of stellar mass loss is considered in great detail. We compute the mass loss of every single population at each epoch after a new stellar population is born. The mass loss of a single population is calculated as follows. We adopt the Scalo initial mass function~\citep{scalo86}. The lifetime of a star with mass ${M}$, $\tau_{M}$, is computed by a broken-power law~\citep{ferreras00}, which is obtained from the data of~\citet{tinsley80} and~\citet{schaller92}.
\begin{eqnarray}
\centering
\tau_{M}({\rm Gyr}) = \left\{ 
\begin{array}{l l}
	9.694({ M/M_{\odot}})^{-2.762} & \quad \mbox{$M<10{M_{\odot}}$}\\
	0.095({M/M_{\odot}})^{-0.764} & \quad \mbox{$M\geq 10{M_{\odot}}$}  \end{array} \right. \
\end{eqnarray}
At the end of its lifetime (after $\tau_{M}$ elapses), a star returns most of its mass into space, leaving small remnants such as a white dwarf, a neutron star or a black hole. The remnant mass of a star with mass $M$, $\omega_{M}$, is suggested by Ferreras $\&$ Silk (2000) as follows:
\begin{eqnarray}
\centering
\frac {\omega_{M}}{M_{\odot}} = \left\{ 
\begin{array}{l l}
	0.1({M/M_{\odot}})+0.45 & \quad \mbox{$M<10{M_{\odot}}$}\\
	1.5 & \mbox{$10{M_{\odot}}\leq M<25{M_{\odot}}$}\\
	0.61({M/M_{\odot}})-13.75 & \quad \mbox{$M\geq 25 {M_{\odot}}$}.  \end{array} \right. \
\end{eqnarray}
In this study, we simply assume that half of the mass loss returns to cold gas components and the rest becomes hot gas.

\subsection{Environmental Effect}

The Chandra X-ray Observatory revealed that massive satellite galaxies in nearby clusters have hot gas components~\citep{sun07,jeltema07}, in contradiction to expectations based on an instantaneous hot gas stripping scenario for satellite galaxies in cluster environments~\citep[e.g.][]{kauffmann99,somerville08}. The old assumption predicted a higher fraction of passive satellites in large halos than observed, known as the satellite overquenching problem~\citep{kimm09}. ~\citet{kimm11} showed that a gradual, rather than instant, and more realistic stripping of the hot gas reservoir relieves the above-mentioned problem to some degree. Therefore, we implemented the gradual hot gas stripping of satellite galaxies in our model by considering tidal stripping~\citep[see][]{kimm11} and ram pressure stripping that uses the prescriptions of~\citet{mccarthy08}, which had been modified for semi-analytic models by~\citet{font08}. However, we are still missing a realistic prescription on cold gas stripping. Empirical evidence for cold gas stripping is clear~\citep{vollmer08,chung08a,chung08b}, and a theoretical study using a semi-analytic approach shows its effect in galaxy evolution~\citep{tecce10}. Thus, it should be considered in our model in due course.

\vspace{4mm}

\subsection{Feedback Processes}
Feedback mechanisms have been introduced into galaxy formation theory to reconcile the discrepancy between galaxy and dark matter halo mass functions. Hydrogen atoms, neutralized at the recombination, may be reionized at later epochs ($z<10$) due to background high-energy photons. This mechanism may suppress the growth of small galaxies~\citep{gnedin00,somerville02,benson02a,benson02b}. Supernova feedback is thought to be effective at disturbing the growth of small galaxies by ejecting cold gas~\citep{white78,dekel86}, and AGN feedback is considered more effective in massive galaxies with a large black hole~\citep{silk98,schawinski06,schawinski07}.

We utilize the reionization prescription of~\citet{benson02b}. The prescription allows an inflow of baryons into a dark halo via accretion of dark matter when $V_{\rm vir} > V_{\rm reionization}$ throughout the age of the Universe, where $V_{\rm reionization}$ is the suppression velocity of reionization . If a halo has a lower value of $V_{\rm vir}$ than the criterion, the inflow of baryons is allowed only at $z > z_{\rm reionization}$, where $z_{\rm reionization}$ is the suppression redshift of reionization. In this study, we adopt $V_{\rm reionization}=30\ {\rm km\ s^{-1}}$ and  $z_{\rm reionization}=8$ following~\citet{benson02b}.

We follow the prescriptions of S08 for supernova feedback. These prescriptions take into account not only the amount of reheated gas but also the fraction of reheated gas blown away from halos. The reheating rate of cold gas due to supernova feedback is formulated as follows:
\begin{eqnarray}
 \dot{m}_{\rm rh}=\epsilon^{\rm SN}_{\rm 0} \left( \frac{\rm 200km\,s^{-1}}{V_{\rm disk}} \right)^{\alpha_{\rm rh}} \dot{m}_{*},
\label{eqn:snf}
\end{eqnarray}
where $\epsilon^{\rm SN}_{\rm 0}$ and $\alpha_{\rm rh}$ are free parameters, $V_{\rm disk}$ is the rotational velocity of a disk, and $\dot{m}_{*}$ is the star formation rate. S08 assumes that the rotational velocity of a disk is the same as the maximum rotational velocity of the DM halo. We calculate the fraction of the reheated gas that has enough kinetic energy to escape from the halo as
\begin{eqnarray}
 f_{\rm eject}=\left[1.0+\left( \frac {V_{\rm vir}}{V_{\rm eject}} \right)^{\alpha_{\rm eject}} \right]^{-1}
\label{eqn:f_eject}
\end{eqnarray}
where $\alpha_{\rm eject}=6$ and $V_{\rm eject} \sim 100-150 {\rm km\, s^{-1}}$. In the case of a satellite halo, a fraction of reheated gas by supernova feedback, $f_{\rm eject}m_{\rm rh}$, is ejected from the satellite and added to the hot gas reservoir of its host halo. On the other hand, it is assumed that the ejected gas from main halos is diffused through inter-cluster medium.  

We take the quasar-mode and radio-mode AGN feedback into account in our model. It has been suggested that the quasar mode is induced by an inflow of cold gas into the central super-massive black hole (SMBH) of the central galaxy during major mergers~\citep{kauffmann00}. The increasing mass of the SMBH via the accretion of cold gas can be expressed as
\begin{eqnarray}
 \Delta m_{\rm BH,Q}=\frac{f_{\rm BH}^{'}m_{\rm cold}}{1+(280{\rm km\, s^{-1}}/V_{\rm vir})^{2}},
\label{eqn:qso}
\end{eqnarray}
where $f_{\rm BH}^{'}$ is the efficiency of gas accretion. In this study, we take the modified parameter proposed by~\citet{croton06}:  $f_{\rm BH}^{'}=f_{\rm BH}(M_{\rm sat}/M_{\rm host})$, where $f_{\rm BH}$ is the original form of the parameter introduced by~\citet{kauffmann00}, $M_{\rm host}$ is the mass of a host galaxy, and $M_{\rm sat}$ is the mass of the host galaxy's merger counterpart. In this study, we assume that the lifetime of the quasar mode, $t_{\rm QSO}$, is 0.2Gyr, as~\citet{martini01} and~\citet{martini03} suggested  $t_{\rm QSO}<0.3$Gyr. Thus, we have $\dot {m}_{\rm BH,Q}=\Delta m_{\rm BH,Q}/t_{\rm QSO}$. It is generally thought that the quasar mode is caused by a high accretion rate of cold gas, resulting in a rapid growth of an SMBH.

\begin{figure}
\centering 
\includegraphics[width=0.45\textwidth]{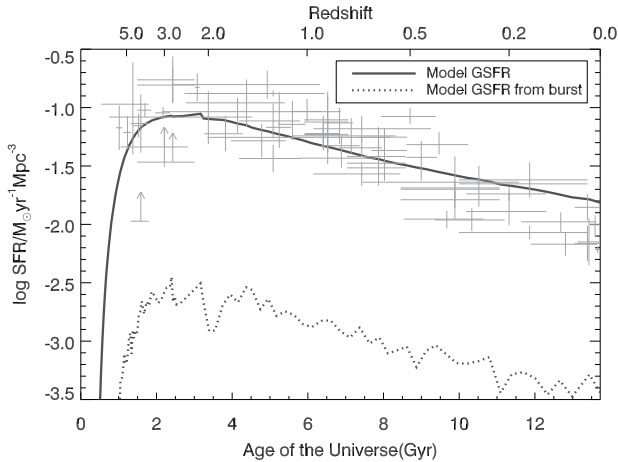}
\caption{The cosmic star formation history. The gray dots with error bars indicate the empirical cosmic star formation history~\citep{panter07}. The solid line shows the cosmic star formation history derived from the fiducial model. The dotted line displays the contribution to the cosmic star formation history from merger-induced starbursts. }
\label{gsfr}
\end{figure}

The radio-mode feedback releases low-Eddington-ratio energy through the accretion of hot gas distributed throughout the halo. Although the energy released from the radio-mode AGN is far less than that from the quasar mode, it is regarded that the radio mode supplies enough energy to the surrounding medium to interrupt gas cooling or to blow away (some of the) cold gas. 
We implement radio-mode feedback into our model following the prescription of~\citet{croton06}:
\begin{eqnarray}
 \dot{ m}_{\rm BH,R}=\kappa_{\rm AGN} \left( \frac{m_{\rm BH}}{10^{8}M_{\odot}} \right) \left( \frac{f_{\rm hot}}{0.1} \right) \left( \frac {V_{\rm vir}}{\rm 200km\, s^{-1}} \right)^{3}
\label{eqn:radio}
\end{eqnarray}
where $\kappa_{\rm AGN}$ is a free parameter with units of ${M_{\odot}\ yr^{-1}}$, $m_{\rm BH}$ is the black hole mass, $f_{\rm hot}$ is the fraction of hot gas with respect to the total halo mass, and $V_{\rm vir}$ is the virial velocity.

We assume that the amount of energy generated by the accretion of gas into the SMBH is given as follows:
\begin{eqnarray}
L_{\rm BH}=\eta \dot{m}_{\rm BH}c^{2}=\eta ( \dot{m}_{\rm BH,Q}+\dot{m}_{\rm BH,R}) c^2,
\label{eqn:agn_energy}
\end{eqnarray}
where $\eta=0.1$ is the standard efficiency of the conversion of rest mass to radiation, and $c$ is the speed of light. The reduced cooling rate of gas is computed by 
\begin{eqnarray}
\dot{m}_{\rm cool}^{'}=\dot{m}_{\rm cool} - \frac {L_{\rm BH}}{0.5V_{\rm vir}^{2}}
\label{eqn:agn_energy}
\end{eqnarray}
where the minimum of $\dot{m}_{\rm cool}^{'}$ is set to be zero.

\begin{figure}
\centering 
\includegraphics[width=0.450\textwidth]{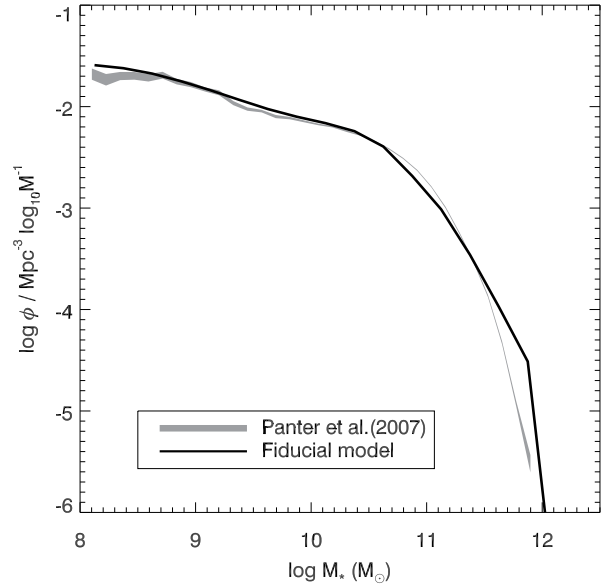}
\caption{The galaxy stellar mass functions in the local Universe derived by~\citet{panter07} from SDSS DR3 (gray shade) and the fiducial model at z=0 (black solid line). The thickness of the empirical data indicates the error range. }
\label{mf}
\end{figure}

\vspace{5mm}

\subsection{Model calibrations}

Our models are based on the conventional techniques and ingredients used in up-to-date semi-analytic models; hence, the output is not particularly noteworthy compared to other successful models. Our models roughly match the global star formation history, galaxy mass functions, black hole mass versus bulge mass relation, etc. Figure~\ref {gsfr} and~\ref{mf} display comparisons of the cosmic star formation history and the galaxy stellar mass functions in the local Universe from empirical data~\citep{panter07} and our fiducial model. While there still is a large room for improvement, we decide to focus on the mass growth histories of massive galaxies.

\vspace{4mm}

\section{Evolution of Galaxy Number Density}

\begin{figure}
\centering 
\includegraphics[width=0.45\textwidth]{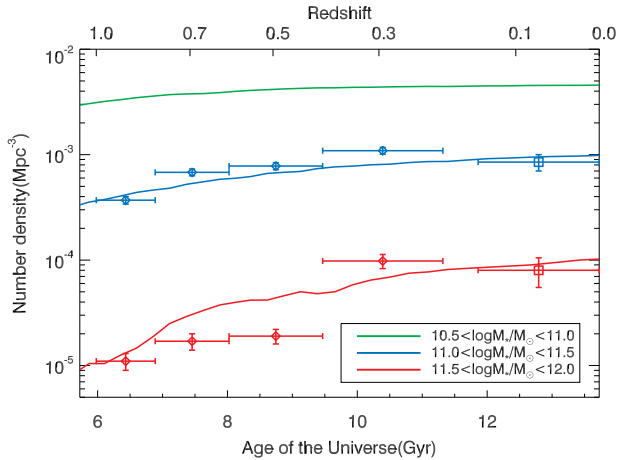}
\caption{ Number density evolution of massive galaxies as a function of the age of the Universe. The red represents the evolution of the most massive galaxies ($\log M/M_{\odot}=11.5-12.0$), the blue indicates that of the next massive group ($\log M/M_{\odot}=11.0-11.5$), and the green indicates that of the third massive group ($\log M/M_{\odot}=10.5-11.0$) at each redshift. The diamonds with error bars come from empirical data in MK10 and the squares with error bars are measurements taken from~\citet{cole01}. The solid lines present the predictions of the semi-analytic model.}
\label{gdensity}
\end{figure}

\begin{figure}
\centering 
\includegraphics[width=0.45\textwidth]{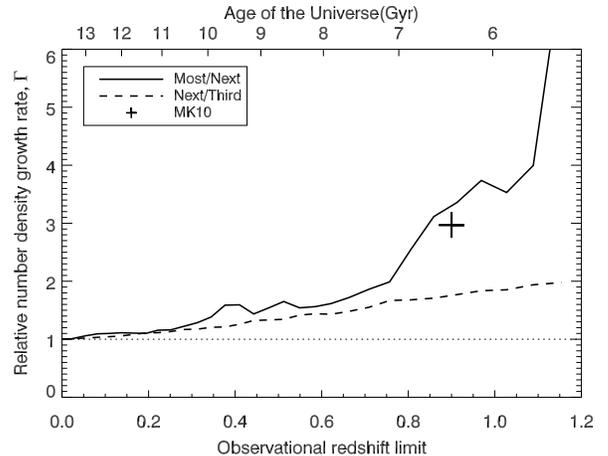}
\caption{ Relative galaxy number density growth rate, $\Gamma$, between z=0 and observational redshift limits. The solid line shows $\Gamma$ of the most/next and the dashed line indicates that of the next/third in our model. The cross is derived from the empirical data in Figure~\ref{gdensity}. }
\label{growth_rate}
\end{figure}

MK10 presented a rapid growth of massive galaxies since $z=1$, using the United Kingdom Infrared Telescope(UKIRT) Infrared Deep Sky Survey (UKIDSS) and the Sloan Digital Sky Survey (SDSS) II Supernova Survey. Figure~\ref{gdensity} shows the number density evolution of the most massive ($\log M/M_{\odot}=11.5-12.0$) and the next massive ($\log M/M_{\odot}=11.0-11.5$) galaxies in the empirical data derived by MK10 and from our fiducial model at each redshift. The empirical data clearly show that the number of the most massive galaxies rapidly increases between $z=1$ and 0, while the next massive group experiences a milder evolution. The reproduction of the data by our fiducial model looks reasonably good. We also present the number density evolution of the third massive group ($\log M/M_{\odot}=10.5-11.0$), which makes our models ``super-$L_{*}$ galaxies''.
We define ``relative number density growth rate'', $\Gamma$, as the ratio of the speeds in number density evolution of the two most massive groups of galaxies as a function of observational limit in redshift, as follows:  
\begin{eqnarray}
\Gamma =\frac {n_{\rm most}({z=0})/n_{\rm most}({z})}{n_{\rm next}({{z=0}})/n_{\rm next}({z})},
\label{eqn:growth_rate}
\end{eqnarray}
where $n_{\rm most}({z=0})$ and  $n_{\rm next}({z=0})$ are the number densities of the most massive and next massive groups at $z=0$, and  $n_{\rm most}(z)$ and  $n_{\rm next}(z)$ are the number densities of the two groups at a redshift, respectively. 
For example, MK10 compared the number density evolution of the two mass groups between redshift 0 and 1, in which case the observational limit is 1 and the relative number density growth rate becomes 3. We present their observations and our models in Figure~\ref{growth_rate}. The MK10 data point should be compared with our model for the most and the next massive galaxy groups (solid line). The model that compares the next massive group of galaxies with $L_*$ galaxies (dashed line) exhibits a similar but milder trend. Observational constraints are still weak but are at least roughly reproduced by the models. The relative growth rate is always greater than 1 in the models, which indicates that {\em the number density of more massive galaxies undergoes a faster evolution than that of less massive groups} in the super-$L_{*}$ range. The number evolution is most dramatic when a comparison is made against the most massive galaxies. This is caused by the fact that the mass bin for the ``most massive'' galaxies has only influx from less massive galaxies, whereas the mass bin of less massive galaxies can have outflux to more massive galaxy bins as well as influx from even less massive galaxy bins.

Primarily motivated by MK10, we focus on the mass growth histories of super-$L_*$ galaxies. We divide the model galaxies into three groups according to $z=0$ mass: Rank 1: $ \log M/M_{\odot}=11.5-12.0$, Rank 2: $\log M/M_{\odot}=11.0-11.5$, and Rank 3: $ \log M/M_{\odot}=10.5-11.0$, where Rank 3 roughly represents $L_{*}$ galaxies. From our simulation volume, we found 49 galaxies in Rank 1, 472 galaxies in Rank 2, and 2,188 galaxies in Rank 3. 

\vspace{5mm}
\section{Evolution of Massive Galaxies in Models}
\subsection{Evolution of Stellar Mass in Galaxies}

\begin{figure}
\centering 
\includegraphics[width=0.45\textwidth]{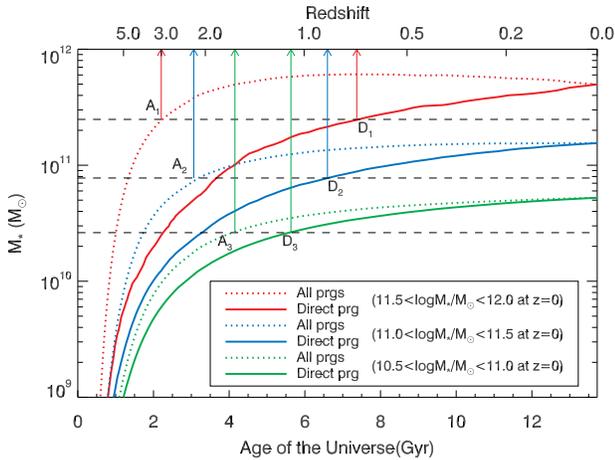}
\caption{ Average mass evolution of galaxies. The red, blue, and green represent three groups of galaxies: $ \log M/M_{\odot}=11.5-12.0$, $\log M/M_{\odot}=11.0-11.5$, and $ \log M/M_{\odot}=10.5-11.0$ at $z=0$, respectively. The solid lines indicate the mean mass evolution of direct progenitors and the dotted lines show the mean mass evolution of all (direct+collateral) progenitors. The black horizontal dashed lines denote half the mean galaxy mass at $z=0$ of the groups. $\rm A_{1}$, $\rm A_{2}$, and $\rm A_{3}$, and the arrows indicate the epochs when the total stellar masses of all progenitors reach half of the final  mass. $\rm D_{1}$, $\rm D_{2}$, and $\rm D_{3}$ with arrows show the epochs at which the direct progenitors of the three groups acquire half of their final  mass.}
\label{mass_evolution}
\end{figure}

In the hierarchical paradigm, a galaxy can have more than one progenitor. Progenitors of a galaxy can be divided into ``direct'' and ``collateral'' progenitors. A direct progenitor is the galaxy in the largest halo when a merger between halos takes place. While there can be numerous progenitors, there is only one direct progenitor at each epoch. Collateral progenitors are all the other galaxies that contribute to the final galaxy. In this concept, to build the evolutionary history of a galaxy, one should consider not only direct progenitors, but also merger counterparts or collateral progenitors.  Figure~\ref{mass_evolution} shows the average mass evolution of the direct progenitors (solid lines) and all (direct and collateral combined) the progenitors (dotted lines) of Rank 1, 2, and 3 galaxies. The mean stellar masses of the three groups at $z=0$ are $4.97 \times 10^{11} M_{\odot}$, $1.55 \times 10^{11} M_{\odot}$, and  $5.25 \times 10^{10} M_{\odot}$,  respectively. Most of the Rank 1 galaxies in our volume are brightest cluster galaxies. 

As all the progenitors merge with each other, the dotted lines and the solid lines finally meet at $z=0$. Stellar mass loss and the scattering of stellar components in satellite galaxies into diffuse stellar components, which takes place when mergers occur, lead to a gradual decrease in the total mass during the evolution. For example, one can see a slight decline in the total stellar mass (red dotted line at the top) after $z\sim 0.5$. This effect is not clearly visible in Rank 2 and 3 galaxies in which star formation is more extended $and$ mergers are less frequent than in Rank 1 galaxies.

It is useful to have a definition of the formation redshift, $z_{\rm f}$. We define it as the redshift at which half of the stellar mass at $z=0$ has been assembled. In the case of Rank 1 galaxies, half of the final mass is achieved at $z_{\rm f,D} \sim0.7$(denoted as $\rm D_{1}$ in Figure~\ref {mass_evolution}) in direct progenitors and at $z_{\rm f,A}\sim3.0$($\rm A_{1}$) when all progenitors are combined. Our models suggest ($z_{\rm f,D}$, $z_{\rm f,A}$) = (0.9, 2.1) for Rank 2, and ($z_{\rm f,D}$, $z_{\rm f,A}$) = (1.1, 1.6) for Rank 3. Models exhibit a monotonic mass dependence of $z_{\rm f,D}$ and $z_{\rm f,A}$ in the sense that, with mass, $z_{\rm f,D}$ decreases while $z_{\rm f,A}$ increases. In other words, the mass of the direct progenitors of a more massive group grows more slowly, while its total mass of all the progenitors is assembled earlier than that of a less massive group. The evolutionary histories of direct progenitors are opposite to the pattern of cosmic downsizing. As~\citet{neistein06} and~\citet{oser10} pointed out, however, if the growth histories of collateral progenitors of galaxies are also considered, downsizing would be a natural outcome of the hierarchical concept of galaxy formation.

\begin{figure}
\centering 
\includegraphics[width=0.45\textwidth]{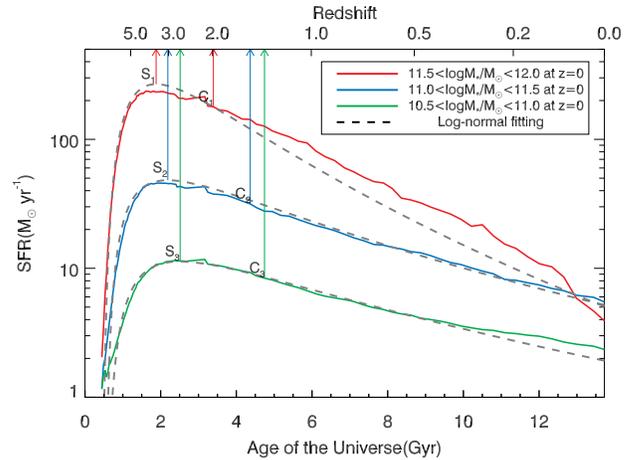}
\caption{Mean star formation histories of all the progenitor galaxies of Rank 1(red), 2(blue), and 3(green). The gray dashed lines denote log-normal fitting. $\rm S_{1}$, $\rm S_{2}$, and $\rm S_{3}$ with arrows indicate the epochs when SFRs peak. $\rm C_{1}$, $\rm C_{2}$, and $\rm C_{3}$ with arrows show the epochs at which the cumulative stellar mass reach half of the total stellar mass born by $z=0$. }
\label{sfr}
\end{figure}

The difference in the growth history between the three groups can be understood in depth through Figure~\ref {sfr}, which presents the evolutionary histories of star formation rates (SFRs).  The mean SFRs contain the star formation histories of both direct and collateral progenitors. We show the best fitting log-normal function to the three SFR curves. The star formation rates of more massive galaxies peak earlier, as marked by S1, S2, and S3 in the figure, and decreases faster than those of less massive groups. The same features were noted in an earlier study by~\citet{delucia06}.

Figure~\ref {sfr} also show the redshift at which half of the total cumulative star formation has occurred in the three groups of galaxies: (C1, C2, C3) = (2.0, 1.5, 1.4) in $z$. The general trend shown by these three values agrees with that of $z_{\rm f,A}$ discussed above. In a sense, it is these values, rather than $z_{\rm f,A}$, that are closer to the general definition of formation redshift. Again, the more massive galaxies are, the earlier they form their stars, consistent with downsizing. In conclusion, {\em the observational finding of downsizing is a result of the hierarchical galaxy formation process, where more massive galaxies have larger stellar ages and smaller assembly ages} ~\citep[see also e.g.][]{delucia07,kaviraj09}. Both the larger stellar ages and the smaller assembly ages can be understood as a result of large-scale effect; that is, in a deeper potential well, progenitor galaxies and their stars form earlier, and many more galaxies participate in galaxy mergers for a long period of time.

\subsection{Origin of Stellar Components}

\begin{figure*}
\centering
\includegraphics[width=1\textwidth]{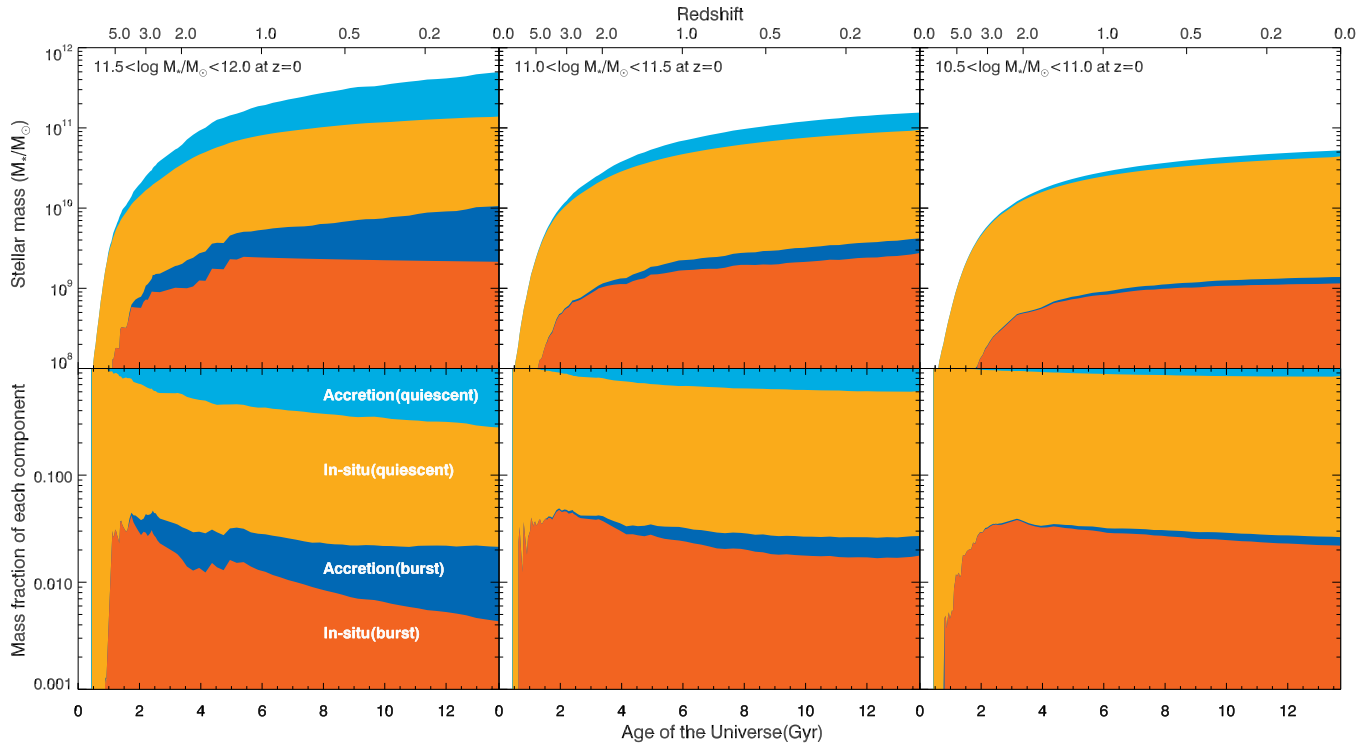}
\caption{ Average mass evolution of stellar components in direct progenitors (upper) and the fraction of each component divided by the mean total stellar mass of  direct progenitors at each redshift (bottom). The left, middle, and right panels show the mean evolutionary histories of the direct progenitors of Rank 1, 2 , and 3 galaxies, respectively. Each color code represents each stellar component as follows: (1) sky blue: merger accretion of stars formed in quiescent mode, (2) orange: {\em in-situ} quiescent star formation, (3) blue: {\em in-situ} starburst, and (4) red: merger accretion of stars formed in burst mode.}  
\label{components}
\end{figure*}

In this section, we investigate how stellar components are assembled into massive galaxies. Stellar components in a galaxy originate either from {\em in-situ} star formation or from ``merger accretion''. Two modes of star formation are considered: ``quiescent'' mode and merger-induced ``starburst''. Stars in a galaxy can therefore have four different origins: (1) {\em in-situ} quiescent star formation, (2) {\em in-situ} starburst, (3) merger accretion of stars formed in quiescent mode, and (4) merger accretion of stars formed in burst mode.

Figure~\ref{components} shows the decomposition of the four channels as a function of time for the three different mass groups. The top and bottom rows show the absolute stellar mass evolution and relative mass fraction evolution, respectively.

Several remarkable features are visible, more easily in the bottom rows. First, {\em in-situ} quiescent star formation is an important channel for the stellar components in massive galaxies. Its fractional contribution is (30, 60, 80)\% in Rank (1, 2, 3) galaxies, respectively. {\em In-situ} quiescent star formation takes place on a galactic disk, and thus, one may wonder why its contribution is so large in these massive, and probably bulge-dominant galaxies. This is because massive bulge-dominant galaxies at $z=0$ have many late-type progenitors, and that is more pronounced in less massive galaxies (Rank 3) than in Rank 1 galaxies. This is a reflection of the progenitor bias discussed earlier~\citep[c.f.][]{franx01,guo08,parry09,kaviraj09}.

Second, the stars formed in burst mode are an extreme minority ($\sim 2$\%) in these massive galaxies. This result is somewhat unexpected. In the hierarchical universe, larger galaxies are generally a product of numerous mergers between galaxies. Mergers, especially between similar-mass galaxies (major mergers), cause a starburst, and so a naive expectation is that larger galaxies would have a large fraction of stars formed in burst mode. However, it is not that simple. Starbursts are usually a result of major mergers, and major mergers are extremely rare once galaxies become massive, that is, at low redshifts. Major mergers between massive galaxies sometimes occur at low redshifts, but they are most often dry, between early-type galaxies without much cold gas~\citep[c.f.][]{vandokkum05,bundy07,bundy09,parry09}. Thus, it has been suggested that the fraction of stars formed in merger-driven burst mode may be less than a few per cent~\citep[e.g.][]{aragon-salamanca98,rodighiero11}.

\begin{figure}
\centering 
\includegraphics[width=0.45\textwidth]{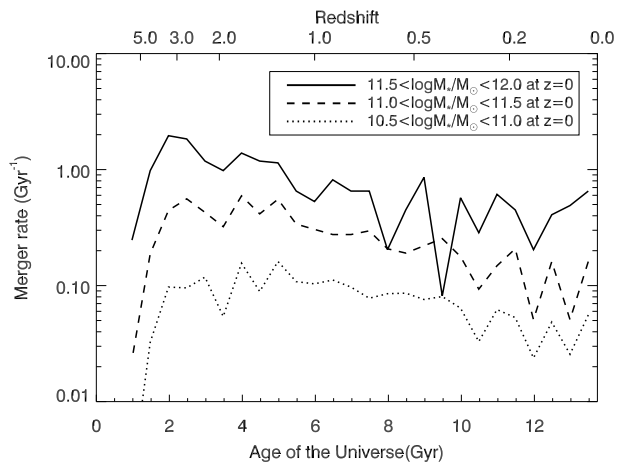}
\caption{Frequency of mergers (per Gyr per galaxy) with a mass ratio $m_2/m_1 \gtrsim 0.1$ experienced by direct progenitors. The solid, dashed, and dotted lines indicate the merger rates of Rank 1, 2, and 3 galaxies, respectively.}
\label{merger_rate}
\end{figure}

Third, merger accretion is the dominant channel in the most massive galaxies. Roughly 70\% of the stellar components in the most massive group (bottom left panel) form outside direct progenitors and get accreted via mergers. This fraction is smaller for Rank 2 (40\%) and Rank 3 (20\%) galaxies, but still substantial. It is interesting to note that much of the mass difference between ranks is attributed to the difference in the amount of merger accretion. For example, roughly 80\% of the mass difference between Ranks 1 and 2 comes from the differences in merger accretion. The value is $\sim 60$\% between Ranks 2 and 3. Massive galaxies are so mainly because they have a substantial amount of {\em in-situ} quiescent star formation, but the {\em most massive} galaxies are so because they have acquired a large amount of mass through {\em merger accretion.} The result is supported by previous studies.~\citet{aragon-salamanca98} confirmed that BCGs have experienced no or negative evolution in luminosity during $0<z<1$ from the K-band Hubble diagram for a sample of BCGs while they have increased their mass by a factor of two to four, depending on cosmological parameters. Thus, it has been understood that merger and accretion may be the most plausible explanation for the evolution. Using a semi-analytic model,~\citet{delucia07} showed that most stellar components in model BCGs are formed in the very early age of the Universe (80 \% at $z\sim3$) in small progenitors and accreted onto BCGs far later (50\% after $z\sim0.5$) via mergers.~\citet{oser10} presented similar results, using numerical simulations. About 80\% of stellar components in simulated massive galaxies ($M_{*}>2.4\times10^{11}M_{\odot}$ at z=0) are formed outside in the early age ($z>3$) and brought into the massive galaxies via mergers and accretion. The massive galaxies double their mass after $z\sim1$. They revealed that the fraction gets smaller in less massive galaxies.~\citet{parry09} also found a similar trend in their investigation on two separate semi-analytic models. They demonstrated that the contribution of mergers to the bulge growth exceeds that of disk instability at $M_{*}>10^{11.5}M_{\odot}$. Considering the fact that such massive galaxies ($M_{*}>10^{11}M_{\odot}$) are likely early type~\citep[e.g.][]{bell03}, it implies that mergers play a more important role in the growth of massive galaxies. It should however be noted that disk instability which is more effective to the smaller late-type galaxy evolution may play a role in such progenitor galaxies of present-day massive early types.

\begin{figure}
\centering 
\includegraphics[width=0.45\textwidth]{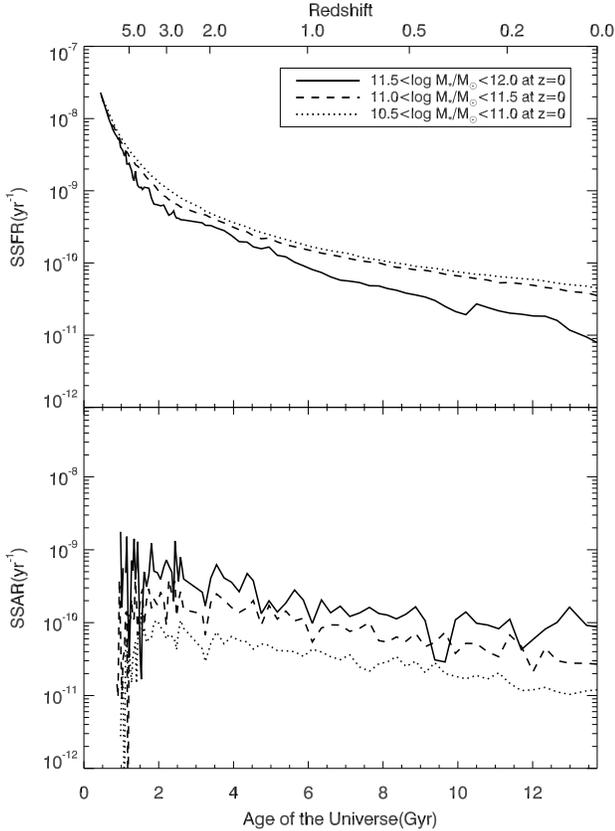}
\caption{The specific star formation rates (upper) and the specific stellar accretion rates via mergers (bottom) of the direct progenitors in the three groups. See the text for their definitions. The solid, dashed, and dotted lines represent Rank 1, 2, and 3 galaxies.}
\label{ssfr_sasr}
\end{figure}

Figure~\ref{merger_rate} shows the merger rate evolution for baryonic mass ratios greater than or equal to 1:10. Although merger rates show stochastic effects, there is a clear decreasing tendency with time, whereas the star formation rates of the three groups drop more sharply, as shown in Figure~\ref{sfr}. In general, more massive galaxies are likely to be involved in galaxy mergers more frequently, so that more massive galaxies have many more stellar components born outside and accreted via mergers, as discussed above. During the whole calculation, Rank 1 galaxies undergo about 9.0 mergers with a mass ratio greater than or equal to 1:10 while those in Ranks 2 and 3 experience 4.0 and 1.0 mergers, respectively. During $ 0<z<1$, Rank 1, 2, and 3 galaxies experience 3.5, 1.8, or 0.6 mergers for the same mass ratio criterion. This explains the higher contribution of merger accretion in more massive galaxies, as illustrated in Figure~\ref{components}.

\subsection{Specific Star Formation Rates and Merger Accretion Rates}

\begin{figure}
\centering 
\includegraphics[width=0.45\textwidth]{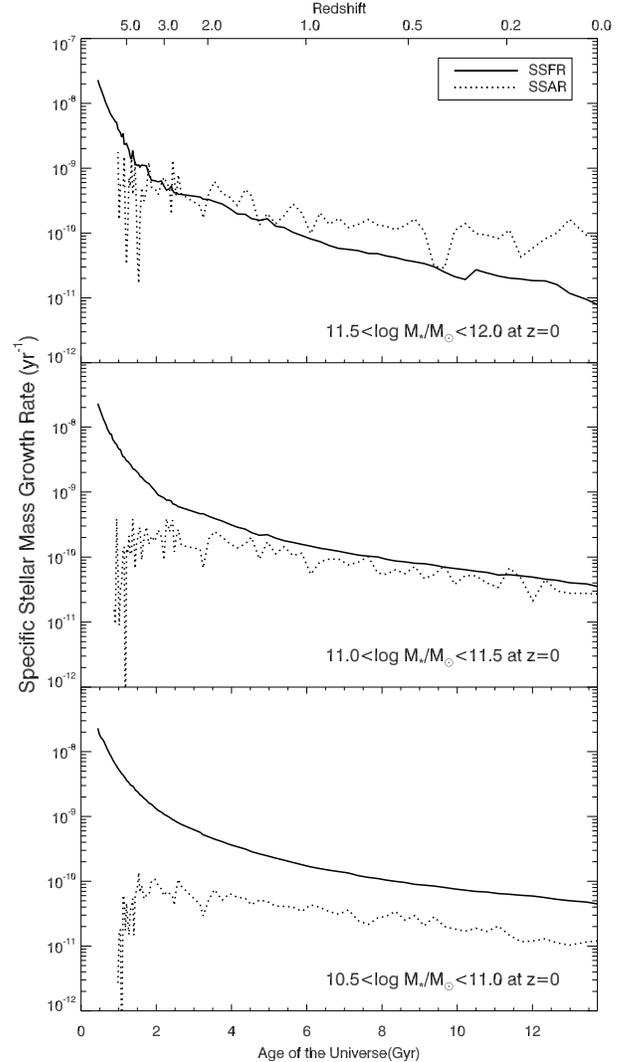}
\caption{Comparison of the mean specific star formation rates (solid lines) and the mean specific stellar accretion rates via mergers (dotted lines) of the direct progenitors of Rank 1(upper),  2 (middle), and 3 (bottom) galaxies.}
\label{dp_ssfr_sasr}
\end{figure}

We found in the previous section that {\em in-situ} star formation and merger accretion were the two most significant channels for the stellar mass growth of massive galaxies. In this section, we scrutinize their time evolution in greater detail.
Specific star formation rates (SSFRs) are normalized growth rates of star formation histories. Likewise, we hereby define the ``specific stellar accretion rate'' (SSAR) to evaluate a normalized growth rate via mergers as follows:
\begin{eqnarray}
{\rm SSAR}= \frac {\Delta M}{ M(t)\Delta t},
\label{eqn:sasr_allprg}
\end{eqnarray}
where $M(t)$ is the mass of a galaxy at an epoch, and $\Delta M$ is an increment of mass by mergers during a time step $\Delta t$. Because we allow diffuse stellar components due to galaxy mergers, mass increment can be expressed as $\Delta M=(1-f_{\rm scatter})M_{\rm *,sat}$, as described in Section 2.1.

Figure~\ref {ssfr_sasr} shows the evolution of the SSFRs (upper) and of the SSARs (bottom) of direct progenitors. In general, more massive galaxies have lower SSFRs, as observations have shown~\citep[e.g.][]{salim07,schiminovich07}, while they have higher SSARs than less massive galaxies. 
Star formation rates are decreased by the depression of gas cooling rates due to an increase in the cooling timescale via the growth of halos, supernova feedback~\citep{white78,dekel86,white91} and/or AGN feedback~\citep{silk98}. Furthermore, the cold gas reservoir of a galaxy could be reduced by feedback processes. Besides, if a galaxy orbits around a more massive galaxy, it becomes redder as it loses its hot gas, which is a source of cold gas, and its cold gas reservoir by tidal and ram pressure stripping~\citep{gunn72,abadi99,quilis00,chung07,tonnesen09,yagi10}. On the other hand, accretion of stellar components via mergers is determined by gravitational interactions between host and subhalos alone; hence, the accretion rate could remain relatively steady as halos continue to fall into other more massive halos over time.

Figure~\ref{dp_ssfr_sasr} displays a comparison of the SSFRs and the SSARs of direct progenitors. As illustrated in Figure~\ref{components}, quiescent star formation dominates the stellar mass growth history in $L_*$ (Rank 3) galaxies (bottom panel). In Rank 2 galaxies, they are comparable to each other most of the time. However, {\em in the most massive (Rank 1) galaxies, merger accretion takes over star formation as the most important channel of stellar mass growth around $z\sim2$} (top panel). 
Our result is qualitatively consistent with that of~\citet{oser10}.

\section{Summary and Discussion}

We have investigated the assembly history of stellar components of massive (super-$L_*$) galaxies, using semi-analytic approaches. Our major results can be summarized as follows.

\begin{itemize}

\item More massive galaxies grow {\em in number} faster than less massive galaxies, as a result of the hierarchical nature of galaxy clustering. This result is consistent with the recent observation of MK10. 

\item The conflict between the predictions from hierarchical models and downsizing {\em is} reconcilable. If we consider only direct progenitors of massive galaxies, our models suggest ``upsizing'' rather than downsizing; that is, the direct progenitors of more massive galaxies grow more slowly. However,  if we consider {\em all} the progenitors, direct and collateral, the combined mass suggests downsizing. Our models suggest that more massive galaxies have older stellar ages but younger assembly ages.

\item Merger-induced ``bursty'' star formation is negligible compared to quiescent disk-mode star formation despite the fact that massive galaxies form through numerous mergers. This is because most of the gas-rich major mergers occur at high redshifts when galaxies are small, and recent major mergers tend to be rare and ``dry''.

\item Merger accretion is a growingly more important channel of stellar mass growth in more massive galaxies. It accounts for 70\% of the final stellar mass in the most massive galaxies in our sample ($\log M/M_{\odot}=11.5-12.0$). It is merger accretion that causes much of the mass difference between massive galaxies. This implies that environments play a central role in the growth of massive galaxies.

\item In the most massive galaxies, which are likely brightest cluster galaxies, merger accretion has remained the most important channel of stellar mass growth ever since $z \sim 2$. 

\end{itemize}

The origin of massive galaxies is a pivotal subject of cosmological paradigms and subsidiary galaxy formation theories. The simplest views based on some pieces of observations, such as downsizing, may favor simplistic formation scenarios, while dynamical models of the universe based on the current cosmology and other statistical aspects of observations, such as galaxy luminosity functions, indicate the other direction. This conflict is at its maximum when it comes to the formation of massive galaxies. The goal of this study is to reconcile the various perspectives and to have an accurate understanding on their formation. In this study, we showed that galaxy models in the hierarchical paradigm provide explanations to seemingly contradicting empirical constraints: the faster growth of the number of more massive galaxies (MK10) and downsizing~\citep{cowie96}. This is encouraging. 

Galaxy formation models, whether hydrodynamic or semi-analytic, are still incomplete in many aspects: it is often claimed that much of their incompleteness is caused by our limited knowledge of baryon physics. The success of understanding {\em massive} galaxy formation on the other hand seems to be more hinged upon our knowledge of large-scale clusterings, and thus dark matter physics. Massive galaxies achieve their grandeur through mergers;  thus, only by a realistic consideration of large-scale clustering information is it possible to accurately reconstruct their formation history.

\section*{acknowledgments}
We thank Taysun Kimm and Sadegh Khochfar for their feedback in the early stage of our code development and Intae Jung for helping us run cosmological volume simulations.
We thank the anonymous referee for a number of comments and suggestions that improved the clarity of the paper. 
We acknowledge the support from the National Research Foundation of Korea through the Center for Galaxy Evolution Research (No. 2010-0027910), Doyak grant (No. 20090078756), and DRC grant.


\vspace{10mm}

\begin{thebibliography}{91}
\expandafter\ifx\csname natexlab\endcsname\relax\def\natexlab#1{#1}\fi

\bibitem[{{Abadi} {et~al.}(1999){Abadi}, {Moore}, \& {Bower}}]{abadi99}
{Abadi}, M.~G., {Moore}, B., \& {Bower}, R.~G. 1999, \mnras, 308, 947

\bibitem[{{Almeida} {et~al.}(2008){Almeida}, {Baugh}, {Wake}, {Lacey},
  {Benson}, {Bower}, \& {Pimbblet}}]{almeida08}
{Almeida}, C., {Baugh}, C.~M., {Wake}, D.~A., {et~al.} 2008, \mnras, 386, 2145

\bibitem[{{Aragon-Salamanca} {et~al.}(1998){Aragon-Salamanca}, {Baugh}, \&
  {Kauffmann}}]{aragon-salamanca98}
{Aragon-Salamanca}, A., {Baugh}, C.~M., \& {Kauffmann}, G. 1998, \mnras, 297,
  427

\bibitem[{{Arnaud} {et~al.}(1992){Arnaud}, {Rothenflug}, {Boulade}, {Vigroux},
  \& {Vangioni-Flam}}]{arnaud92}
{Arnaud}, M., {Rothenflug}, R., {Boulade}, O., {Vigroux}, L., \&
  {Vangioni-Flam}, E. 1992, \aap, 254, 49

\bibitem[{{Arp}(1966)}]{arp66}
{Arp}, H. 1966, \apjs, 14, 1

\bibitem[{{Aubert} {et~al.}(2004){Aubert}, {Pichon}, \& {Colombi}}]{aubert04}
{Aubert}, D., {Pichon}, C., \& {Colombi}, S. 2004, \mnras, 352, 376

\bibitem[{{Battin}(1987)}]{battin87}
{Battin}, R.~H. 1987, {An introduction to the mathematics and methods of
  astrodynamics.} (New York: AIAA)

\bibitem[{{Baugh} {et~al.}(1996){Baugh}, {Cole}, \& {Frenk}}]{baugh96}
{Baugh}, C.~M., {Cole}, S., \& {Frenk}, C.~S. 1996, \mnras, 283, 1361

\bibitem[{{Bell} {et~al.}(2003){Bell}, {McIntosh}, {Katz}, \&
  {Weinberg}}]{bell03}
{Bell}, E.~F., {McIntosh}, D.~H., {Katz}, N., \& {Weinberg}, M.~D. 2003, \apjs,
  149, 289

\bibitem[{{Benson} {et~al.}(2002{\natexlab{a}}){Benson}, {Frenk}, {Lacey},
  {Baugh}, \& {Cole}}]{benson02a}
{Benson}, A.~J., {Frenk}, C.~S., {Lacey}, C.~G., {Baugh}, C.~M., \& {Cole}, S.
  2002{\natexlab{a}}, \mnras, 333, 177

\bibitem[{{Benson} {et~al.}(2002{\natexlab{b}}){Benson}, {Lacey}, {Baugh},
  {Cole}, \& {Frenk}}]{benson02b}
{Benson}, A.~J., {Lacey}, C.~G., {Baugh}, C.~M., {Cole}, S., \& {Frenk}, C.~S.
  2002{\natexlab{b}}, \mnras, 333, 156

\bibitem[{{Binney} \& {Tremaine}(2008)}]{binney08}
{Binney}, J., \& {Tremaine}, S. 2008, {Galactic Dynamics: Second Edition}
  (Princeton University Press)

\bibitem[{{Borne} {et~al.}(2000){Borne}, {Bushouse}, {Lucas}, \&
  {Colina}}]{borne00}
{Borne}, K.~D., {Bushouse}, H., {Lucas}, R.~A., \& {Colina}, L. 2000, \apjl,
  529, L77

\bibitem[{{Bower} {et~al.}(1992){Bower}, {Lucey}, \& {Ellis}}]{bower92}
{Bower}, R.~G., {Lucey}, J.~R., \& {Ellis}, R.~S. 1992, \mnras, 254, 601

\bibitem[{{Bundy} {et~al.}(2009){Bundy}, {Fukugita}, {Ellis}, {Targett},
  {Belli}, \& {Kodama}}]{bundy09}
{Bundy}, K., {Fukugita}, M., {Ellis}, R.~S., {et~al.} 2009, \apj, 697, 1369

\bibitem[{{Bundy} {et~al.}(2007){Bundy}, {Treu}, \& {Ellis}}]{bundy07}
{Bundy}, K., {Treu}, T., \& {Ellis}, R.~S. 2007, \apjl, 665, L5

\bibitem[{{Chung} {et~al.}(2008{\natexlab{a}}){Chung}, {van Gorkom}, {Crowl},
  {Kenney}, \& {Vollmer}}]{chung08a}
{Chung}, A., {van Gorkom}, J.~H., {Crowl}, H., {Kenney}, J.~D.~P., \&
  {Vollmer}, B. 2008{\natexlab{a}}, in Astronomical Society of the Pacific
  Conference Series, Vol. 395, Frontiers of Astrophysics: A Celebration of
  NRAO's 50th Anniversary, ed. A.~H. {Bridle}, J.~J. {Condon}, \& G.~C. {Hunt},
  364

\bibitem[{{Chung} {et~al.}(2008{\natexlab{b}}){Chung}, {van Gorkom}, {Kenney},
  {Crowl}, {Vollmer}, \& {Schiminovich}}]{chung08b}
{Chung}, A., {van Gorkom}, J.~H., {Kenney}, J., {et~al.} 2008{\natexlab{b}}, in
  Astronomical Society of the Pacific Conference Series, Vol. 396, Formation
  and Evolution of Galaxy Disks, ed. J.~G. {Funes} \& E.~M. {Corsini}, 127

\bibitem[{{Chung} {et~al.}(2007){Chung}, {van Gorkom}, {Kenney}, \&
  {Vollmer}}]{chung07}
{Chung}, A., {van Gorkom}, J.~H., {Kenney}, J.~D.~P., \& {Vollmer}, B. 2007,
  \apjl, 659, L115

\bibitem[{{Cimatti} {et~al.}(2004){Cimatti}, {Daddi}, {Renzini}, {Cassata},
  {Vanzella}, {Pozzetti}, {Cristiani}, {Fontana}, {Rodighiero}, {Mignoli}, \&
  {Zamorani}}]{cimatti04}
{Cimatti}, A., {Daddi}, E., {Renzini}, A., {et~al.} 2004, \nat, 430, 184

\bibitem[{{Cole} {et~al.}(2001){Cole}, {Norberg}, {Baugh}, {Frenk},
  {Bland-Hawthorn}, {Bridges}, {Cannon}, {Colless}, {Collins}, {Couch},
  {Cross}, {Dalton}, {De Propris}, {Driver}, {Efstathiou}, {Ellis},
  {Glazebrook}, {Jackson}, {Lahav}, {Lewis}, {Lumsden}, {Maddox}, {Madgwick},
  {Peacock}, {Peterson}, {Sutherland}, \& {Taylor}}]{cole01}
{Cole}, S., {Norberg}, P., {Baugh}, C.~M., {et~al.} 2001, \mnras, 326, 255

\bibitem[{{Cowie} {et~al.}(1996){Cowie}, {Songaila}, {Hu}, \&
  {Cohen}}]{cowie96}
{Cowie}, L.~L., {Songaila}, A., {Hu}, E.~M., \& {Cohen}, J.~G. 1996, \aj, 112,
  839

\bibitem[{{Cox} {et~al.}(2008){Cox}, {Jonsson}, {Somerville}, {Primack}, \&
  {Dekel}}]{cox08}
{Cox}, T.~J., {Jonsson}, P., {Somerville}, R.~S., {Primack}, J.~R., \& {Dekel},
  A. 2008, \mnras, 384, 386

\bibitem[{{Croton} {et~al.}(2006){Croton}, {Springel}, {White}, {De Lucia},
  {Frenk}, {Gao}, {Jenkins}, {Kauffmann}, {Navarro}, \& {Yoshida}}]{croton06}
{Croton}, D.~J., {Springel}, V., {White}, S.~D.~M., {et~al.} 2006, \mnras, 365,
  11

\bibitem[{{De Lucia} \& {Blaizot}(2007)}]{delucia07}
{De Lucia}, G., \& {Blaizot}, J. 2007, \mnras, 375, 2

\bibitem[{{De Lucia} {et~al.}(2006){De Lucia}, {Springel}, {White}, {Croton},
  \& {Kauffmann}}]{delucia06}
{De Lucia}, G., {Springel}, V., {White}, S.~D.~M., {Croton}, D., \&
  {Kauffmann}, G. 2006, \mnras, 366, 499

\bibitem[{{Dekel} \& {Silk}(1986)}]{dekel86}
{Dekel}, A., \& {Silk}, J. 1986, \apj, 303, 39

\bibitem[{{Feldmeier} {et~al.}(2002){Feldmeier}, {Mihos}, {Morrison}, {Rodney},
  \& {Harding}}]{feldmeier02}
{Feldmeier}, J.~J., {Mihos}, J.~C., {Morrison}, H.~L., {Rodney}, S.~A., \&
  {Harding}, P. 2002, \apj, 575, 779

\bibitem[{{Ferreras} \& {Silk}(2000)}]{ferreras00}
{Ferreras}, I., \& {Silk}, J. 2000, \apj, 532, 193

\bibitem[{{Font} {et~al.}(2008){Font}, {Bower}, {McCarthy}, {Benson}, {Frenk},
  {Helly}, {Lacey}, {Baugh}, \& {Cole}}]{font08}
{Font}, A.~S., {Bower}, R.~G., {McCarthy}, I.~G., {et~al.} 2008, \mnras, 389,
  1619

\bibitem[{{Franx} \& {van Dokkum}(2001)}]{franx01}
{Franx}, M., \& {van Dokkum}, P.~G. 2001, in Astronomical Society of the
  Pacific Conference Series, Vol. 230, Galaxy Disks and Disk Galaxies, ed.
  J.~G. {Funes} \& E.~M. {Corsini}, 581--588

\bibitem[{{Glazebrook} {et~al.}(2004){Glazebrook}, {Abraham}, {McCarthy},
  {Savaglio}, {Chen}, {Crampton}, {Murowinski}, {J{\o}rgensen}, {Roth}, {Hook},
  {Marzke}, \& {Carlberg}}]{glazebrook04}
{Glazebrook}, K., {Abraham}, R.~G., {McCarthy}, P.~J., {et~al.} 2004, \nat,
  430, 181

\bibitem[{{Gnedin}(2000)}]{gnedin00}
{Gnedin}, N.~Y. 2000, \apj, 542, 535

\bibitem[{{Gonzalez} {et~al.}(2005){Gonzalez}, {Zabludoff}, \&
  {Zaritsky}}]{gonzalez05}
{Gonzalez}, A.~H., {Zabludoff}, A.~I., \& {Zaritsky}, D. 2005, \apj, 618, 195

\bibitem[{{Gunn} \& {Gott}(1972)}]{gunn72}
{Gunn}, J.~E., \& {Gott}, III, J.~R. 1972, \apj, 176, 1

\bibitem[{{Guo} \& {White}(2008)}]{guo08}
{Guo}, Q., \& {White}, S.~D.~M. 2008, \mnras, 384, 2

\bibitem[{{Jarosik} {et~al.}(2011){Jarosik}, {Bennett}, {Dunkley}, {Gold},
  {Greason}, {Halpern}, {Hill}, {Hinshaw}, {Kogut}, {Komatsu}, {Larson},
  {Limon}, {Meyer}, {Nolta}, {Odegard}, {Page}, {Smith}, {Spergel}, {Tucker},
  {Weiland}, {Wollack}, \& {Wright}}]{jarosik11}
{Jarosik}, N., {Bennett}, C.~L., {Dunkley}, J., {et~al.} 2011, \apjs, 192, 14

\bibitem[{{Jeltema} {et~al.}(2007){Jeltema}, {Mulchaey}, {Lubin}, \&
  {Fassnacht}}]{jeltema07}
{Jeltema}, T.~E., {Mulchaey}, J.~S., {Lubin}, L.~M., \& {Fassnacht}, C.~D.
  2007, \apj, 658, 865

\bibitem[{{Jiang} {et~al.}(2008){Jiang}, {Jing}, {Faltenbacher}, {Lin}, \&
  {Li}}]{jiang08}
{Jiang}, C.~Y., {Jing}, Y.~P., {Faltenbacher}, A., {Lin}, W.~P., \& {Li}, C.
  2008, \apj, 675, 1095

\bibitem[{{Kauffmann}(1996)}]{kauffmann96}
{Kauffmann}, G. 1996, \mnras, 281, 487

\bibitem[{{Kauffmann} {et~al.}(1999){Kauffmann}, {Colberg}, {Diaferio}, \&
  {White}}]{kauffmann99}
{Kauffmann}, G., {Colberg}, J.~M., {Diaferio}, A., \& {White}, S.~D.~M. 1999,
  \mnras, 303, 188

\bibitem[{{Kauffmann} \& {Haehnelt}(2000)}]{kauffmann00}
{Kauffmann}, G., \& {Haehnelt}, M. 2000, \mnras, 311, 576

\bibitem[{{Kauffmann} {et~al.}(1993){Kauffmann}, {White}, \&
  {Guiderdoni}}]{kauffmann93}
{Kauffmann}, G., {White}, S.~D.~M., \& {Guiderdoni}, B. 1993, \mnras, 264, 201

\bibitem[{{Kaviraj}(2010)}]{kaviraj10}
{Kaviraj}, S. 2010, \mnras, 408, 170

\bibitem[{{Kaviraj} {et~al.}(2009){Kaviraj}, {Devriendt}, {Ferreras}, {Yi}, \&
  {Silk}}]{kaviraj09}
{Kaviraj}, S., {Devriendt}, J.~E.~G., {Ferreras}, I., {Yi}, S.~K., \& {Silk},
  J. 2009, \aap, 503, 445

\bibitem[{{Kaviraj} {et~al.}(2007){Kaviraj}, {Schawinski}, {Devriendt},
  {Ferreras}, {Khochfar}, {Yoon}, {Yi}, {Deharveng}, {Boselli}, {Barlow},
  {Conrow}, {Forster}, {Friedman}, {Martin}, {Morrissey}, {Neff},
  {Schiminovich}, {Seibert}, {Small}, {Wyder}, {Bianchi}, {Donas}, {Heckman},
  {Lee}, {Madore}, {Milliard}, {Rich}, \& {Szalay}}]{kaviraj07}
{Kaviraj}, S., {Schawinski}, K., {Devriendt}, J.~E.~G., {et~al.} 2007, \apjs,
  173, 619

\bibitem[{{Kimm} {et~al.}(2011){Kimm}, {Yi}, \& {Khochfar}}]{kimm11}
{Kimm}, T., {Yi}, S.~K., \& {Khochfar}, S. 2011, \apj, 729, 11

\bibitem[{{Kimm} {et~al.}(2009){Kimm}, {Somerville}, {Yi}, {van den Bosch},
  {Salim}, {Fontanot}, {Monaco}, {Mo}, {Pasquali}, {Rich}, \& {Yang}}]{kimm09}
{Kimm}, T., {Somerville}, R.~S., {Yi}, S.~K., {et~al.} 2009, \mnras, 394, 1131

\bibitem[{{Kodama} \& {Arimoto}(1997)}]{kodama97}
{Kodama}, T., \& {Arimoto}, N. 1997, \aap, 320, 41

\bibitem[{{Lackner} {et~al.}(2012){Lackner}, {Cen}, {Ostriker}, \&
  {Joung}}]{lackner12}
{Lackner}, C.~N., {Cen}, R., {Ostriker}, J.~P., \& {Joung}, M.~R. 2012, \mnras,
  3426

\bibitem[{{Macci{\`o}} {et~al.}(2007){Macci{\`o}}, {Dutton}, {van den Bosch},
  {Moore}, {Potter}, \& {Stadel}}]{maccio07}
{Macci{\`o}}, A.~V., {Dutton}, A.~A., {van den Bosch}, F.~C., {et~al.} 2007,
  \mnras, 378, 55

\bibitem[{{Martini} \& {Schneider}(2003)}]{martini03}
{Martini}, P., \& {Schneider}, D.~P. 2003, \apjl, 597, L109

\bibitem[{{Martini} \& {Weinberg}(2001)}]{martini01}
{Martini}, P., \& {Weinberg}, D.~H. 2001, \apj, 547, 12

\bibitem[{{Matsuoka} \& {Kawara}(2010)}]{matsuoka10}
{Matsuoka}, Y., \& {Kawara}, K. 2010, \mnras, 405, 100

\bibitem[{{McCarthy} {et~al.}(2008){McCarthy}, {Frenk}, {Font}, {Lacey},
  {Bower}, {Mitchell}, {Balogh}, \& {Theuns}}]{mccarthy08}
{McCarthy}, I.~G., {Frenk}, C.~S., {Font}, A.~S., {et~al.} 2008, \mnras, 383,
  593

\bibitem[{{Monaco} {et~al.}(2006){Monaco}, {Murante}, {Borgani}, \&
  {Fontanot}}]{monaco06}
{Monaco}, P., {Murante}, G., {Borgani}, S., \& {Fontanot}, F. 2006, \apjl, 652,
  L89

\bibitem[{{Moster} {et~al.}(2010){Moster}, {Somerville}, {Maulbetsch}, {van den
  Bosch}, {Macci{\`o}}, {Naab}, \& {Oser}}]{moster10}
{Moster}, B.~P., {Somerville}, R.~S., {Maulbetsch}, C., {et~al.} 2010, \apj,
  710, 903

\bibitem[{{Murante} {et~al.}(2004){Murante}, {Arnaboldi}, {Gerhard}, {Borgani},
  {Cheng}, {Diaferio}, {Dolag}, {Moscardini}, {Tormen}, {Tornatore}, \&
  {Tozzi}}]{murante04}
{Murante}, G., {Arnaboldi}, M., {Gerhard}, O., {et~al.} 2004, \apjl, 607, L83

\bibitem[{{Navarro} {et~al.}(1996){Navarro}, {Frenk}, \& {White}}]{navarro96}
{Navarro}, J.~F., {Frenk}, C.~S., \& {White}, S.~D.~M. 1996, \apj, 462, 563

\bibitem[{{Neistein} {et~al.}(2006){Neistein}, {van den Bosch}, \&
  {Dekel}}]{neistein06}
{Neistein}, E., {van den Bosch}, F.~C., \& {Dekel}, A. 2006, \mnras, 372, 933

\bibitem[{{Oser} {et~al.}(2010){Oser}, {Ostriker}, {Naab}, {Johansson}, \&
  {Burkert}}]{oser10}
{Oser}, L., {Ostriker}, J.~P., {Naab}, T., {Johansson}, P.~H., \& {Burkert}, A.
  2010, \apj, 725, 2312

\bibitem[{{Panter} {et~al.}(2007){Panter}, {Jimenez}, {Heavens}, \&
  {Charlot}}]{panter07}
{Panter}, B., {Jimenez}, R., {Heavens}, A.~F., \& {Charlot}, S. 2007, \mnras,
  378, 1550

\bibitem[{{Parry} {et~al.}(2009){Parry}, {Eke}, \& {Frenk}}]{parry09}
{Parry}, O.~H., {Eke}, V.~R., \& {Frenk}, C.~S. 2009, \mnras, 396, 1972

\bibitem[{{Quilis} {et~al.}(2000){Quilis}, {Moore}, \& {Bower}}]{quilis00}
{Quilis}, V., {Moore}, B., \& {Bower}, R. 2000, Science, 288, 1617

\bibitem[{{Rodighiero} {et~al.}(2011){Rodighiero}, {Daddi}, {Baronchelli},
  {Cimatti}, {Renzini}, {Aussel}, {Popesso}, {Lutz}, {Andreani}, {Berta},
  {Cava}, {Elbaz}, {Feltre}, {Fontana}, {F{\"o}rster Schreiber},
  {Franceschini}, {Genzel}, {Grazian}, {Gruppioni}, {Ilbert}, {Le Floch},
  {Magdis}, {Magliocchetti}, {Magnelli}, {Maiolino}, {McCracken}, {Nordon},
  {Poglitsch}, {Santini}, {Pozzi}, {Riguccini}, {Tacconi}, {Wuyts}, \&
  {Zamorani}}]{rodighiero11}
{Rodighiero}, G., {Daddi}, E., {Baronchelli}, I., {et~al.} 2011, \apjl, 739,
  L40

\bibitem[{{Salim} {et~al.}(2007){Salim}, {Rich}, {Charlot}, {Brinchmann},
  {Johnson}, {Schiminovich}, {Seibert}, {Mallery}, {Heckman}, {Forster},
  {Friedman}, {Martin}, {Morrissey}, {Neff}, {Small}, {Wyder}, {Bianchi},
  {Donas}, {Lee}, {Madore}, {Milliard}, {Szalay}, {Welsh}, \& {Yi}}]{salim07}
{Salim}, S., {Rich}, R.~M., {Charlot}, S., {et~al.} 2007, \apjs, 173, 267

\bibitem[{{Scalo}(1986)}]{scalo86}
{Scalo}, J.~M. 1986, in IAU Symposium, Vol. 116, Luminous Stars and
  Associations in Galaxies, ed. C.~W.~H. {De Loore}, A.~J. {Willis}, \&
  P.~{Laskarides}, 451--466

\bibitem[{{Schaller} {et~al.}(1992){Schaller}, {Schaerer}, {Meynet}, \&
  {Maeder}}]{schaller92}
{Schaller}, G., {Schaerer}, D., {Meynet}, G., \& {Maeder}, A. 1992, \aaps, 96,
  269

\bibitem[{{Schawinski} {et~al.}(2007){Schawinski}, {Thomas}, {Sarzi},
  {Maraston}, {Kaviraj}, {Joo}, {Yi}, \& {Silk}}]{schawinski07}
{Schawinski}, K., {Thomas}, D., {Sarzi}, M., {et~al.} 2007, \mnras, 382, 1415

\bibitem[{{Schawinski} {et~al.}(2006){Schawinski}, {Khochfar}, {Kaviraj}, {Yi},
  {Boselli}, {Barlow}, {Conrow}, {Forster}, {Friedman}, {Martin}, {Morrissey},
  {Neff}, {Schiminovich}, {Seibert}, {Small}, {Wyder}, {Bianchi}, {Donas},
  {Heckman}, {Lee}, {Madore}, {Milliard}, {Rich}, \& {Szalay}}]{schawinski06}
{Schawinski}, K., {Khochfar}, S., {Kaviraj}, S., {et~al.} 2006, \nat, 442, 888

\bibitem[{{Schiminovich} {et~al.}(2007){Schiminovich}, {Wyder}, {Martin},
  {Johnson}, {Salim}, {Seibert}, {Treyer}, {Budav{\'a}ri}, {Hoopes},
  {Zamojski}, {Barlow}, {Forster}, {Friedman}, {Morrissey}, {Neff}, {Small},
  {Bianchi}, {Donas}, {Heckman}, {Lee}, {Madore}, {Milliard}, {Rich}, {Szalay},
  {Welsh}, \& {Yi}}]{schiminovich07}
{Schiminovich}, D., {Wyder}, T.~K., {Martin}, D.~C., {et~al.} 2007, \apjs, 173,
  315

\bibitem[{{Schweizer} \& {Seitzer}(1988)}]{schweizer88}
{Schweizer}, F., \& {Seitzer}, P. 1988, \apj, 328, 88

\bibitem[{{Sheen} {et~al.}(2012){Sheen}, {Yi}, {Ree}, \& {Lee}}]{sheen12}
{Sheen}, Y.-K., {Yi}, S.~K., {Ree}, C.~H., \& {Lee}, J. 2012, ArXiv e-prints

\bibitem[{{Silk} \& {Rees}(1998)}]{silk98}
{Silk}, J., \& {Rees}, M.~J. 1998, \aap, 331, L1

\bibitem[{{Somerville}(2002)}]{somerville02}
{Somerville}, R.~S. 2002, \apjl, 572, L23

\bibitem[{{Somerville} {et~al.}(2008){Somerville}, {Hopkins}, {Cox},
  {Robertson}, \& {Hernquist}}]{somerville08}
{Somerville}, R.~S., {Hopkins}, P.~F., {Cox}, T.~J., {Robertson}, B.~E., \&
  {Hernquist}, L. 2008, \mnras, 391, 481

\bibitem[{{Spergel} {et~al.}(2007){Spergel}, {Bean}, {Dor{\'e}}, {Nolta},
  {Bennett}, {Dunkley}, {Hinshaw}, {Jarosik}, {Komatsu}, {Page}, {Peiris},
  {Verde}, {Halpern}, {Hill}, {Kogut}, {Limon}, {Meyer}, {Odegard}, {Tucker},
  {Weiland}, {Wollack}, \& {Wright}}]{spergel07}
{Spergel}, D.~N., {Bean}, R., {Dor{\'e}}, O., {et~al.} 2007, \apjs, 170, 377

\bibitem[{{Springel}(2005)}]{springel05}
{Springel}, V. 2005, \mnras, 364, 1105

\bibitem[{{Springel} {et~al.}(2006){Springel}, {Frenk}, \&
  {White}}]{springel06}
{Springel}, V., {Frenk}, C.~S., \& {White}, S.~D.~M. 2006, \nat, 440, 1137

\bibitem[{{Springel} {et~al.}(2001){Springel}, {White}, {Tormen}, \&
  {Kauffmann}}]{springel01}
{Springel}, V., {White}, S.~D.~M., {Tormen}, G., \& {Kauffmann}, G. 2001,
  \mnras, 328, 726

\bibitem[{{Sun} {et~al.}(2007){Sun}, {Jones}, {Forman}, {Vikhlinin}, {Donahue},
  \& {Voit}}]{sun07}
{Sun}, M., {Jones}, C., {Forman}, W., {et~al.} 2007, \apj, 657, 197

\bibitem[{{Sutherland} \& {Dopita}(1993)}]{sutherland93}
{Sutherland}, R.~S., \& {Dopita}, M.~A. 1993, \apjs, 88, 253

\bibitem[{{Tecce} {et~al.}(2010){Tecce}, {Cora}, {Tissera}, {Abadi}, \&
  {Lagos}}]{tecce10}
{Tecce}, T.~E., {Cora}, S.~A., {Tissera}, P.~B., {Abadi}, M.~G., \& {Lagos},
  C.~D.~P. 2010, \mnras, 408, 2008

\bibitem[{{Tinsley}(1980)}]{tinsley80}
{Tinsley}, B.~M. 1980, \fcp, 5, 287

\bibitem[{{Tonnesen} \& {Bryan}(2009)}]{tonnesen09}
{Tonnesen}, S., \& {Bryan}, G.~L. 2009, \apj, 694, 789

\bibitem[{{Tweed} {et~al.}(2009){Tweed}, {Devriendt}, {Blaizot}, {Colombi}, \&
  {Slyz}}]{tweed09}
{Tweed}, D., {Devriendt}, J., {Blaizot}, J., {Colombi}, S., \& {Slyz}, A. 2009,
  \aap, 506, 647

\bibitem[{{van Dokkum}(2005)}]{vandokkum05}
{van Dokkum}, P.~G. 2005, \aj, 130, 2647

\bibitem[{{Vollmer} {et~al.}(2008){Vollmer}, {Braine}, {Pappalardo}, \&
  {Hily-Blant}}]{vollmer08}
{Vollmer}, B., {Braine}, J., {Pappalardo}, C., \& {Hily-Blant}, P. 2008, \aap,
  491, 455

\bibitem[{{White} \& {Frenk}(1991)}]{white91}
{White}, S.~D.~M., \& {Frenk}, C.~S. 1991, \apj, 379, 52

\bibitem[{{White} \& {Rees}(1978)}]{white78}
{White}, S.~D.~M., \& {Rees}, M.~J. 1978, \mnras, 183, 341

\bibitem[{{Woods} \& {Geller}(2007)}]{woods07}
{Woods}, D.~F., \& {Geller}, M.~J. 2007, \aj, 134, 527

\bibitem[{{Yagi} {et~al.}(2010){Yagi}, {Yoshida}, {Komiyama}, {Kashikawa},
  {Furusawa}, {Okamura}, {Graham}, {Miller}, {Carter}, {Mobasher}, \&
  {Jogee}}]{yagi10}
{Yagi}, M., {Yoshida}, M., {Komiyama}, Y., {et~al.} 2010, \aj, 140, 1814

\bibitem[{{Zibetti} {et~al.}(2005){Zibetti}, {M{\'e}nard}, {Nestor}, \&
  {Turnshek}}]{zibetti05}
{Zibetti}, S., {M{\'e}nard}, B., {Nestor}, D., \& {Turnshek}, D. 2005, \apjl,
  631, L105

\end{thebibliography}
\end{document}